\documentclass{PoS}

\title{Continuous smearing of Wilson Loops.}

\ShortTitle{Continuous smearing of Wilson Loops.}

\author{Robert Lohmayer\thanks{Research supported in 
part by the DOE, grant nr. DE-FG02-01ER41165}\\
        Rutgers University, Department of Physics and 
Astronomy, Piscataway, NJ 08854\\
        E-mail: \email{Lohamyer@physics.rutgers.edu}}

\author{\speaker{Herbert Neuberger}%
         \thanks{Research supported in part by the DOE, 
grant nr. DE-FG02-01ER41165}\\
        Rutgers University, Department of Physics and 
Astronomy, Piscataway, NJ 08854\\
        E-mail: \email{neuberg@physics.rutgers.edu}}

\abstract{Continuum smearing was introduced in section 
4.1 of JHEP03, 064 (2006) as a meaningful continuum 
analogue of the well known set of lattice techniques 
by the same name. Here we apply continuous smearing 
in continuous space-time to Wilson loops in order to 
clarify what it does in the context
of field theory and also in the context of the loop 
calculus of the Makeenko-Migdal equation.}

\FullConference{XXIX International Symposium on Lattice Field Theory \\
                 July 10--16, 2011\\
                 Squaw Valley, Lake Tahoe, California}

\begin{document}

\section{What is smearing?}

In standard form, 
smearing is a map $S[\phi_0 (x)]=\phi_s(x)$ of general 
structure $\partial_s \phi_s (x) = F[\phi_s (x)]$ 
where $F[.]$ has some locality property and
an initial condition is supplied at $s=0$. 
$\phi_0 (x)$ denotes a set of fields defined
on a smooth manifold and $x$ is a point on that manifold.
For us, $x$ is a point in space-time, and the fields take
values in some other space, for example the space of a 
compact Lie Group. 
The smearing parameter $s\ge 0$
is a dimensionful quantity, a length squared in the length units of $x$. 
$S$ acts as a UV filter on the $x$-dependence, 
by an amount $s$.
Smearing is extremely widely used in science and technology 
under different names of similar meaning. For a mathematical
overview of linear smoothing we refer to~\cite{shoen}.

In this write-up we shall focus on smearing 
in the context of gauge field theory.

\section{A brief history of smearing in gauge field theory}

Smearing is a technique in gauge theory whose scope
has evolved over time. Many contributions have been made over the years 
and its present form is an outgrowth of these. 
We only mention the few instances
where the scope of smearing was extended 
in a qualitative manner.

In classical Yang Mills theory smearing enters under the name
``gradient flow''~\cite{donal_atiya}. Here the fields
are classical and smooth and so is the flow parameter, which  
we denoted by $s$ (for smearing). 

In lattice field
theory smearing seems to have been employed first by the APE
group~\cite{ape}. The new ingredient here is that
smearing is completely discretized: the fields live on a
lattice (no longer in the continuum) and the variable
$s$ is also discretized into smearing steps. Many variants
of smearing have appeared over 
the years within the same scope,
optimizing on various features. We only mention 
some of the various terms that were used, in addition to
``smearing'': ``cooling'', ``smoothing'', ``fattening'', 
``UV filtering'', ``Gaussian filtering'', ``HYP''.
The smeared fields were viewed as an intermediate
step to constructing operators which ultimately were thought of 
as complicated expressions in terms of the original  
lattice fields, designed to facilitate extracting some 
specific physical quantity. In this context smearing was viewed 
as a tweaking of lattice artifacts with the objective of shortening
simulation time or enhancing the speed of convergence to the continuum limit. 
Relatively to the classical situation 
encountered in mathematics, the new ingredient was that the
input into the smearing process became a quantum field 
which was fluctuating and non-smooth; it was UV cutoff
dependent. One did only a finite number of discrete smearing
steps and strict lattice locality was preserved.
Consequently, the standard way of taking the continuum 
limit still applied. Two essential 
parameters characterized a 
smearing procedure: a step size and a number of steps. It was
physically clear however, that only their product should really matter. 

The scope of smearing was further enlarged by studying the 
perturbative expansion of smeared lattice operators, including
their dependence on the smearing details~\cite{bernard}. 
This work clarified how the total amount of
smearing (step size times number of steps) becomes a continuous
parameter of dimension length squared. Thus, the smearing
flow parameter became smooth again, just like in the classical
case, but space-time remained discrete and the UV cutoff
remained explicit. Perturbation theory for fat links is described in
section 15 of~\cite{capitani}.

The last time the scope of smearing was enlarged seems to be
in~\cite{nn}. In section 4.1 in that paper the explicit
UV cutoff was dropped also, going back to the classical 
situation in the sense that smearing was
now a procedure completely defined 
within continuum gauge field theory. The main 
observation was that the renormalization of 
smeared observables presented none of the 
new divergences associated with composite operators in the 
traditional, unsmeared, setting: 

Standard renormalization produces finite Greens functions
for products of elementary fields at distinct space-time
points. By ``elementary fields'' one means the fields that
appear explicitly in the Lagrangian. As is well
known, renormalization needs to be employed again when
one wants to consider Greens functions which include local
operators composed of elementary fields and their derivatives
at the same space-time point. Depending on the
complexity of the operator, a large number of arbitrary
finite parts will be required to end up with well defined
expressions. This issue is particularly
relevant in a gauge theory, where all physical operators are
composite since gauge invariance is better viewed 
as a redundancy of the elementary-field representation and the
concept of an ``elementary field'' is not physical. 
For example, the classic choice for 
a complete set of physical operators is 
the collection of all Wilson loops. Their renormalization
is well understood and the various 
standard ways to define finite correlation functions 
of Wilson loops ultimately obscure the basic
fact that parallel transport round a closed curve 
ought to have a monodromy represented by
an element of a compact group. 

Continuous
smearing eliminates these problems at the expense of 
introducing a smearing parameter. The classic renormalization
of the elementary fields is enough to ensure finiteness and 
uniqueness of all smeared Wilson loops, so long as
the smearing parameter is nonzero. 

With this it seems that
the circle has been closed, and classical smearing 
has been fully extended to the quantum level. 
Considering a product of smeared composite operators, each at
its own positive smearing parameter,
the well known structures of the operator product
expansion will emerge in a limit 
where some of the smearing parameters are taken to zero.
In the case of a Wilson loop, one can lift the loop to 
$R^4\times R_+$, with a continuously varying smearing parameter.

Just as the correlation 
functions of elementary fields in ordinary field theory are defined (up to 
well understood ambiguities) by the Lagrangian, 
one may consider smeared correlation 
functions of all smeared fields, whether composite or not, 
also as well defined: 
One is free to supplement the classical Lagrangian by
any quantization procedure one chooses and then
calculate the same smeared correlation functions one
would in any other procedure. A natural 
framework in the context of smearing is Parisi-Wu
quantization, which employs the Langevin equation. 
Quantum fluctuations are induced in this case 
by a white noise (a regularization can be introduced
by controlling the high momentum modes of the noise), 
which abstracts the interaction with a heat bath. 
Without smearing, one waits until thermal equilibration 
and then extracts correlations. Smearing means that after
equilibration one takes the heat bath temperature abruptly 
down to zero, and then waits some more, specifically, an amount of 
Langevin-time equal to
the smearing parameter $s$, before extracting 
correlation functions. During the cooling interval of Langevin-time extent
$s$, the noise is set to zero. 

\section{Smearing of gauge theories}

We now get very specific to the gauge theory case. 
Smearing limits the resolution at which we describe physics, by
a small length scale $\sqrt{s}$. This length scale is finite in
physical units, say, the inverse of the square root of the string tension.
The string tension is defined in the absence of smearing. 
The limitation on resolution must preserve gauge invariance.
In equations this means
\begin{eqnarray*} S[A^\omega ]=\{S[A]\}^\omega, \forall \{ A,\omega\},\end{eqnarray*}
where $\omega$ is the infinitesimal local gauge transformation 
applied to the connection $A$. 
In perturbation theory gauge invariance can be replaced by BRST invariance. 
This also works on the lattice~\cite{brst}. The BRST formalism does not really
need a Lagrangian: the equations of motion are sufficient. Of course, they
have to be gauge covariant. This means that smearing does not have to be
introduced at the Lagrangian level; we shall use an equation of motion that does
not follow from a Lagrangian.  The lowest order partial differential equation
obeying all symmetries with smearing included is~\cite{nn}:
\begin{eqnarray*}
F_{\mu,s} = D_{\nu} F_{\mu,\nu}.
\end{eqnarray*}
In terms of number of  derivatives, the above 
equation is a minimal form of smearing, which indicates 
universality. 
The equation is taken on $R^4\times R_+$. $R^4$ can 
be replaced by any compact manifold.
The first order nature in terms of $s$ requires a single
boundary condition which is taken as 
\begin{eqnarray*}
A_\mu(x;0) = B_\mu (x),
\end{eqnarray*}
where the boundary field $B$ is the ordinary quantum field on $R^4$.
Thus, the quantum fluctuations are in full strength at $s=0$ but 
as $s$ increases, the limitation on the resolution limits their impact
on the smeared observables. 

The noncompact nature of the $R_+$ component strongly
suggests a standard gauge choice: $A_s=0$. In this gauge the gauge field
smearing equation is brought to standard form. 
In terms of the general definition of smearing given earlier 
our ``minimal'' equation of motion set the functional $F[.]$ 
to $-\frac{\delta {\bf A} [A]}{\delta A_\mu (x)}$, 
where ${\bf A}$ is the classical Yang Mills action. 
One advantage of this
choice for $F[.]$ is that the original Lagrangian 
of the theory also defines the smearing
procedure. Hence, any regularization that is defined at the Lagrangian
level (like, for example, a lattice or a Pauli-Villars regularization) 
naturally produces a smearing procedure.

\section{Smeared Wilson loops}

 Introduce a loop operator 
$\Psi ({\cal C},x,s)={\cal P} e^{ig_0\oint_{{\cal C},x}^x A_\mu(x,s) dx_\mu }$.
The Wilson loops are $W_r[{\cal C},s] =\langle \chi_r [\Psi({\cal C},x,s)]\rangle$
with $r$ an irreducible representation and $s>0$. We use the 
normalization $\chi_r ({\bf 1})\equiv 1$. 
$g_0$ is the bare
coupling constant.  ${\cal C}$ is a smooth closed curve and $x$ is a point on it.
$A_\mu(x,s)$ is the smeared gauge field defining a traceless 
hermitian matrix valued one form
on $R^4$, smoothly parameterized by $s$. 

For any $s>0$, after the introduction of the standard counter terms,  
the calculation of $W_r$ produces no new UV divergences. 
Consequently, $\Psi$ can be thought of as a 
fluctuating $SU(N)$ matrix, strictly obeying the 
constraints $\Psi^\dagger \Psi=1$ and $\det \Psi=1$.
This remains true beyond perturbation theory. 

Other regularizations of $W_r$ tend to violate the bound $|W_r|\le 1$ which
would follow from $\Psi$ being a fluctuating unitary matrix. 
This happens because the
ordinary Coulomb term enters with a plus sign in the exponent and the negative perimeter
divergence is thrown out. Smearing, on the other hand, 
regulates away the singularity of the 
Coulomb potential at short distance and replaces the negative divergent perimeter term
by a negative term with a $\frac{1}{\sqrt{s}}$ dependence. So far, smeared Wilson loops
seem to be the single option available to maintain the $SU(N)$-matrix character of the 
quantum variable $\Psi$ simultaneously in perturbative and 
non-perturbative quantum gauge theory 
Moreover, $W_r[{\cal C},s]$ is a re-parameterization and zig-zag invariant
Stokes functional of ${\cal C}$. 
Later, we shall argue that if there exists, in some sense,
a dual string representation of $W_r[{\cal C},s]$ for $s=0$, there
should exist one also for $s>0$.

We first restrict our attention to curves that are continuous, have a continuous
tangent and do not self-intersect.
Define couplings $g^2_w$ by $\log\left [ W_r[{\cal C},s]\right ]=-g^2_w [{\cal C},s,r]
C_2(r) f_{\rm tree} [{\cal C},s]$, where $f_{\rm tree}$ 
is defined by tree level perturbation theory, when
$g^2_w\to g_0^2$. 
We parameterize the curves by an overall scale $l$, the perimeter of the curve, 
and a set of shape variables $\zeta_\alpha$.
${\cal C}$ now refers to all curves that are translations and rigid
rotations of the same curve. The set $\zeta_\alpha$ is invariant under dilatations. 
The entire dependence on dilatations resides in $l$.
The function $f_{\rm tree}$ depends only on the $\zeta_\alpha$ 
and on $\frac{l}{\sqrt{s}}$.  For $l,\sqrt{s}\ll \Lambda_N^{-1}$, 
fixed $\zeta_\alpha$ and $\frac{l}{\sqrt{s}}$, we can 
calculate the perturbative running of $g^2_w$ as a function of $l$.  

$f_{\rm tree}$ is a functional of ${\cal C}$ already defined in abelian Yang Mills theory.
Assuming a fixed amount of smearing, one has a short distance expansion for $l\to 0$
and a long distance one for $l\to\infty$:
The long distance expansion, for 
$l\to \infty$, gives:
\begin{eqnarray*} f_{\rm tree} [{\cal C},s]= a_0[{\cal C}] 
\frac{l}{\sqrt{s}} + a_1 [{\cal C}] + 
a_2[{\cal C}]\frac{\sqrt{s}}{l}+...
\end{eqnarray*}
At tree level the theory is conformal, and therefore the
real expansion parameter is the ratio $\frac{l}{\sqrt{s}}$, so
large loops of a given shape correspond to some fixed size loop of the
same shape, but at high resolution. 

$a_0 [{\cal C}]$ equals $\frac{1}{8\pi\sqrt{2\pi}}$, 
providing the coefficient of what would become
the perimeter divergent term when $s\to 0$ in the abelian case. 
$a_1[{\cal C}]$ only depends on the $\zeta_\alpha$ and is proportional to 
the ``number of photons'' first introduced (with different motivation) 
by L. Stodolsky~\cite{stodo}:
\begin{eqnarray*}
a_1[{\cal C}]=\frac{1}{4\pi^2} \oint dx_\mu \oint dy_\nu 
\left [ \frac{\delta_{\mu\nu}}{(x-y)^2}-\frac{(x-y)_\mu (x-y)_\nu}{(x-y)^4}\right ].
\end{eqnarray*}
The integrand in the above expression is the photon propagator in a particular gauge,
but, going from that gauge to Feynman gauge for example, involves a gauge function that
is singular on the loop. This is how the above expression, 
which is finite, becomes 
formally gauge equivalent to an expression that has a perimeter divergence.
The same manipulation in the smeared case is valid, as the singularity is smeared
away. Thus, smearing provides yet another way to arrive 
at Stodolsky's formula, by explicitly removing the term  
$a_0[{\cal C}] \frac{l}{\sqrt{s}}$ and taking $s\to 0$ afterwards. 
Moreover, since $g^2_w [{\cal C},s,r]$ is well defined in the 
fully interacting nonabelian quantum field theory, 
the long distance expansion of $g^2_w$ provides a possible 
extension of Stodolsky's photon number to the
non-abelian case.

Similarly, the short distance expansion, 
$l\to 0$, looks like
\begin{eqnarray*}  f_{\rm tree} [{\cal C},s] = b_0 [{\cal C}]\frac{l^4}{s^2} + 
b_1[{\cal C}]\frac{l^6}{s^3}+...
\end{eqnarray*}
Here, 
\begin{eqnarray*}
b_0[{\cal C}]= \frac{1}{2^{9} \pi^2} \frac{\sum_{\mu\nu} \sigma_{\mu,\nu}^2}{l^4},
\end{eqnarray*}
where
$\sigma_{\mu,\nu}=\frac{1}{2}\oint (dx_\mu x_\nu - dx_\nu x_\mu)$ is the 
enclosed area of the projection of ${\cal C}$ on the $\mu,\nu$ plane. 
In this case the expectation value defining $g_w^2$ is ${\rm Tr}_f
\langle F^2_{\mu\nu}(x,s)\rangle$. 
At one higher order we get terms of the form 
$\langle (DF)^2_{\mu\nu}(x,s)\rangle$, etc.

The structure of this short distance
expansion extends from the tree level all the way
beyond perturbation theory. A short distance expansion of Wilson 
loops was discussed in~\cite{shif}. There the associated condensates 
were described employing a physically sensible, albeit 
imprecise, concept of separation
between perturbative and nonperturbative contributions. 
Smearing provides one possible disambiguation of this description.  

If we allow isolated discontinuities in the tangent to
the curve there are logarithmic
corner divergences when $s\to 0$, locally 
associated with each discontinuity in the tangent. 

For specific simple loop shapes one can calculate $f_{\rm tree}[{\cal C}]$
explicitly. 
For example, consider a circle of radius $R$. We have, with $\rho=\frac{2s}{R^2}$,
\begin{eqnarray*}
f_{\rm tree}({\rm circle}) = -\frac{1}{4}+\frac{e^{-\frac{1}{2\rho}}}{8\rho}
\left [ (1+2\rho ) I_0 \left (\frac{1}{2\rho}\right )+I_1 \left ( \frac{1}{2\rho}\right )
\right ],
\end{eqnarray*}
where the $I_n(.)$ are modified Bessel functions. 

For a rectangular $T\times R$ loop, we get a more complicated expression in terms of an integral
with an error function inside the integrand. In an asymptotic expansion for large loops
one recovers the would be linear and logarithmic divergences. Beyond that,
one has domination by the Coulomb term, linear
in $\frac{R}{T}+\frac{T}{R}$. Such a term also appears in an effective
string representation expected to be valid for large rectangular loops.  There 
it no longer is multiplied by a coupling constant, but rather by a universal
number. This gives an indication how one might be able to
match perturbative to string expressions. 

For numerical purposes, $f_{\rm tree}$ has also been calculated on the lattice with a 
single plaquette Wilson action.

\section{Smearing in loop calculus}

As is well known, one expects Yang Mills theory with gauge group $SU(N)$
to be related to some string theory with a string coupling going to
zero as $N\to\infty$. Thus, at least in the large-$N$ limit,
one does not need a completely well defined string theory;  
just a well defined generalized two dimensional sigma-model would do, if 
it were the right one. To be sure, the observable we are interested in 
in this two dimensional field theory is a complicated loop functional,
so even if somebody handed us the right sigma-model we still would 
face a difficult problem. At $N=\infty$ the set of field-theoretical 
Schwinger Dyson equations for Wilson loops closes 
on loops in the fundamental representation, 
producing the Makeenko-Migdal~\cite{makeenko} loop equation.
It should be solved by the loop functional coming from the sigma-model. The loop
equation  is elegant only at a formal, unregularized level. In 
the smeared setup, it is a property of the initial condition, 
expressed in terms of the boundary fields $B$. The attractiveness of
the equation stems from the fact that all the Yang Mills
structure has disappeared, being replaced by 
a partial differential equation on loop space. 
More specifically, there appears a loop Laplacian, whose definition
is relatively well understood. The solution to the equation should be a member 
of a specific subset of all possible loop functionals: it must be
of ``Stokes'' type, be reparameterization invariant and obey a zig-zag symmetry. 

The above properties hold for smeared Wilson loops. One would therefore like
to be able to write down some loop equation directly for smeared loops.
This would provide an opening to make the required defining property
of the dual field theoretic -- string representations concrete, as it 
would deal with well defined objects. 

Although this was not mentioned in~\cite{nn},
the original motivation for choosing smearing as a way to handle 
composite operators 
was that it could be represented in loop space 
with the help of the same loop Laplacian which appears in the 
Makeenko-Migdal equation: 
\begin{eqnarray*}
\frac {\partial W_r [{\cal C},s]}{\partial s}=\oint d\sigma \frac{\delta^2 W_r[{\cal C},s]}
{\delta x_\mu^2 (\sigma)}.
\end{eqnarray*}
The original Makeenko-Migdal equation places a constraint on the 
initial data, at $s=0$. To what degree this fully determines the 
initial data takes us back to the unsolved problem of
finding a finite, elegant version of the Makeenko-Migdal
equation. Nevertheless, we can say that, if indeed a string route
is found which makes sense of the original Makeenko-Migdal
equation, this route would also make sense of a smeared version
of the solution to these equation 
and therefore smearing should extend to the string theory. 
Alternatively, another view of the Makeenko-Migdal equation in
the smeared context is as an asymptotic condition at $s\to 0$ 
on a solution of the loop-space heat equation describing smearing.

\section{Hamiltonian smearing}

Smearing as defined above will obscure constraints on correlation 
functions which represent Minkowski space unitarity. It is 
possible to avoid this by smearing only in the space directions.

\end{document}